\documentclass[useAMS,usenatbib]{mn2e}
\usepackage{graphicx}
\usepackage{natbib}
\usepackage{bm}
\usepackage{amsmath}
\usepackage{amsfonts}
\usepackage{color}
\usepackage{arydshln}
\usepackage{tabularx}
\usepackage{colortbl}
\usepackage{hhline}
\usepackage{siunitx}

\usepackage{array}
\usepackage{multirow}

\newcommand\MyBox[2]{
  \fbox{\lower0.75cm
    \vbox to 1.0cm{\vfil
      \hbox to 2.0cm{\hfil\parbox{1.4cm}{#1\\#2}\hfil}
      \vfil}%
  }%
}

\usepackage{placeins} 

\newcommand{\url}{{url\;}}

\title[Physics-aware Neural Networks]{Self-supervised Learning with Physics-aware Neural Networks I: Galaxy Model Fitting}
\author[Aragon-Calvo M.A. et al.]{M.A. Aragon-Calvo$^{1}$ \thanks{E-mail:maragon@astro.unam.mx} \\
$^{1}$Instituto de Astronom\'{i}a, UNAM, Apdo. Postal 106, Ensenada 22800, B.C., M\'{e}xico\\}
\begin{document}

\date{}

\pagerange{\pageref{firstpage}--\pageref{lastpage}} \pubyear{2002}
\maketitle
\label{firstpage}

\begin{abstract}

Estimating the parameters of a model describing a set of observations using a neural network is in general solved in a supervised way. 
In cases when we do not have access to the model's true parameters this approach can not be applied. Standard unsupervised learning techniques on the other hand, do not produce meaningful or \textit{semantic} representations that can be associated to the model's parameters.
Here we introduce a self-supervised hybrid network that combines traditional neural network elements with analytic or numerical  models which represent a physical process to be learned by the system. Self-supervised learning is achieved by generating an internal representation equivalent to the parameters of the  physical model. This semantic representation is used to evaluate the model and compare it to the input data during training.
The \textit{Semantic Autoencoder} architecture described here shares the robustness of neural networks while including an explicit model of the data, learns in an unsupervised way and estimates, by construction, parameters with direct physical interpretation. As an illustrative application we perform unsupervised learning for 2D model fitting of exponential light profiles.

\end{abstract}

\section{Introduction}\label{sec:intro}

Fitting a model to a given set of observations is a basic paradigm in science. Model fitting allows us to estimate the value of the parameters defining the model and also serves as a tool for testing its validity. Traditionally, this task has been approached as an optimization problem where we wish to minimize the error between the model $G$ and a set of observations ${\bm x}$ such as done in regression techniques \citep{Legendre}. Model fitting can be a challenging task as often the result is highly sensitive to the set of initial parameters, noise and artifacts in the data. 

In recent years, new techniques developed in the field of Machine Learning (ML) have been successfully applied to problems where the  complexity of the data prevented the use of traditional techniques. One of the fields that has seen the largest advances is image processing, where Deep Neural Network (DNN) architectures have now reached and even surpassed human performance \citep{LeCun15, Dieleman15}. At the core of the ML revolution is the ability of DNN to generate an internal hierarchical representation of the patterns in the data. However, they do so at the cost of a lack of transparency.  Typical DNN systems contain millions of trainable parameters with no meaningful interpretation, effectively making them black boxes. This lack of transparency in the inner workings of DNN still limits their full application in quantitative sciences. Currently, DNN are mainly used in astronomy as workhorses for classification, segmentation and regression tasks \citep{Dieleman15, Kim17, Tuccillo18, Sanchez18}. However, their use as tools for insight extraction of physical processes remains poorly exploited. The work presented here aims to be one step forward in this direction by explicitly introducing physical models in a regular neural net architecture.

\begin{figure*}
  \centering
  \includegraphics[width=0.8\textwidth,angle=0.0]{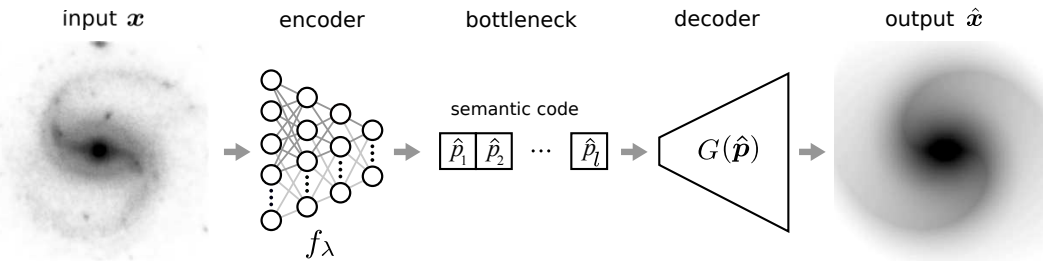}
  \caption{Semantic autoencoder. The input image ${\bm x}$ represents a measurement taken from some process $\Psi$. The encoder $f_{\lambda}$ consists of a standard deep neural network that maps the input into the semantic parameters $ \hat{{\bm p}}$. The decoder model $G$ evaluates $\hat{{\bm p}}$ producing the reconstructed output $\hat{{\bm x}}$. The decoder $G$ is a model that can be analytically or numerically solved and has not trainable parameters. Note that the actual output of interest of this architecture is the semantic code at the bottleneck, not the output image.}\label{fig:autoencoder}
\end{figure*}

%


%
\subsection{Autoencoders}

Autoencoders belong to a family of neural networks that are trained to copy its input to its output. While replicating the input is a trivial task the utility of autoencoders stems form their ability to map the high-dimensional input into a low-dimensional compressed internal representation \citep{LeCun87,Bourlard88,Hinton06}. 

A standard autoencoder is composed of an encoder and a decoder joined at a bottleneck (see Fig. \ref{fig:autoencoder}). 
The encoder is a neural network $f_{\lambda}$ (described by a set of trainable parameters $\lambda$ consisting of weights and biases) that maps the input 
${\bm x} \in \mathbb{R}^m$ 
into a code 
${\bm c}  = f_{\lambda} ({\bm x}) \in \mathbb{R}^l$. 
The code is then fed into a decoder neural network $g_{\lambda}$ that ``decompresses'' it back to input space producing a reconstruction of the input: $\hat{\bm x} = g_{\lambda} ({\bm c}) = g_{\lambda} (f_{\lambda} ({\bm x}) ) \in \mathbb{R}^m$. Learning the optimal $\lambda$ is performed by minimizing the loss function $L ({\bm x} , \hat{\bm x} )$ chosen to quantify the difference between the input and the reconstructed output.

In order for autoencoders to produce a compressed representation they are, by construction,  undercomplete  i.e. the dimensionality of the code is lower than that of the input ($m > l$). One consequence of this is that the reconstruction will not be an exact copy of the input data. In fact, it can be desirable that the reconstructed output is a simplified version of the input, meaning that only the relevant aspects of the input image are encoded \citep{Vincent10}.

Autoencoders are attractive architectures for data encoding due to their ability to create a compressed representation at their bottleneck. This representation is in general not semantically meaningful (although it has been shown that sparcity can lead to semantical encoding for simple physical models \citep{Iten18}). In general sparsity is a required but not sufficient condition for semantic encoding. This is also the case of other dimensionality reduction techniches such as Principal Component Analysis, for which the linear autoencoder produces equivalent encodings \citep{Bourlard88}.

%
\subsection{Semantic autoencoders}

The lack of semantic encoding in regular autoencoders is a consequence, in part, of their generality as both the encoder and decoder are based on universal trainable systems \citep{Cybenko89}. One way of enforcing semantic encoding would be to impose meaningful constraints on the autoencoder.  Consider a physical process $\Psi$ from which we take a set of measurements ${\bm x}$. The process $\Psi$ can be described or approximated by a (mathematical) model $G$ expressed as a function of a set of parameters ${\bm p}$:

\begin{equation}
\Psi({\bm p}) \to {\bm x } \approx G ( {\bm p} ).
\end{equation}

\noindent We now feed ${\bm x}$ to an autoencoder that will encode ${\bm x}$  into the parameters $\hat{{\bm p}} = f_{\lambda}({\bm x})$ at its bottleneck and then decode it back to produce the reconstructed input  $\hat{\bm x} = g_{\lambda} (\hat{{\bm p}})$. 
If we somehow  force $\hat{{\bm p}}$  and ${\bm p}$ to be semantically equivalent (${\bm p} \Leftrightarrow \hat{{\bm p}}$) then the decoder $g_{\lambda}$ will effectively learn to mimic the behavior of the process $\Psi$.

 We can then replace the decoder $g_{\lambda}$ by the model $G$ as they are functionally equivalent ($g_{\lambda} \Leftrightarrow G$). The constraints imposed on the autoencoder by $G$ force the encoder $f_{\lambda}$ to generate a semantic encoding of the input. Note that the model $G$ can be any function such as a modeling equation, a numerical code like an N-body solver, or even an actual physical system such as a mechanical arm. Semantic autoencoders do not require to be provided the ``true'' code ${\bm p}$ during training (we may not even have access to the code). Instead the rely on the validity of the model $G$ in order to generate reconstructed observations ${\hat{\bm x}}$ and learn in a self-supervised way.

The {\it semantic autoencoder} described here is a system combining trainable with non-trainable elements. In this approach we consider neural networks as special cases of a larger family of differentiable functions with trainable or non trainable parameters that can learn via gradient descent \cite{Wengert64,Bischof96}.

\begin{figure*}
  \centering
  \includegraphics[width=0.99\textwidth,angle=0.0]{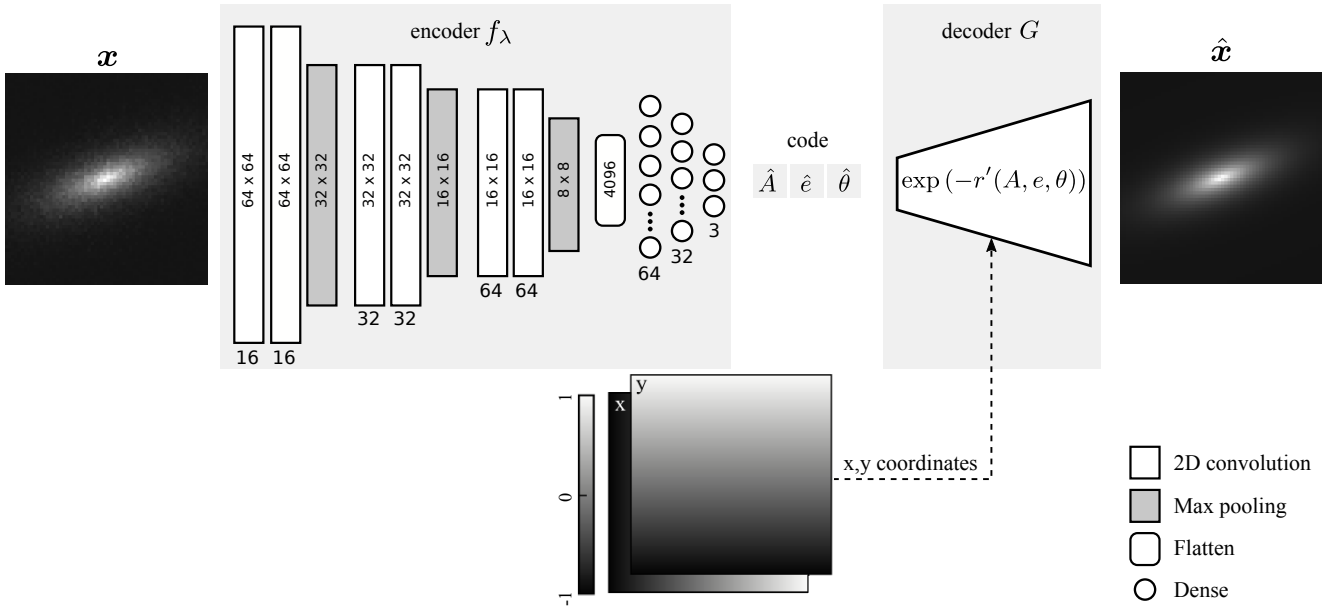}
  \caption{Implementation of a semantic autoencoder trained to estimate the parameters $A, e$ and $\theta$ defining a 2D exponential light profile. The input (noisy) image $\hat{{\bm x}}$ is feed to an encoder $f_{\lambda}$ consisting of a deep convolutional neural network and a dense neural network. At the bottleneck we obtain the parameters $\hat{A}, \hat{e}, \hat{\theta}$ which are then used to evaluate the model $G$ (here corresponding to eq. \ref{eq:exp}) in the decoder, yielding the reconstructed output image. The numbers under the layers indicate the number of neurons/kernel in the layer. The horizontal numbers indicate the activation image size in pixels. During training we compare, via the loss function, the input and output images. Note the auxiliary input consisting of a two-channel 2D array containing the $x$ and $y$ coordinates used to expand the model's parameters to a 2D image.}\label{fig:implementation}
\end{figure*}

\section{Self-supervised exponential profile fitting}

In this section we present an illustrative application of a semantic autoencoder trained to estimate the parameters defining an exponential profile commonly used to describe the light distribution of late-type galaxies \citep{Sersic63,Peng02, Simard11, Barden12}. Model fitting can be addressed as a standard supervised problem  providing both the input image and the model's parameters \citep{Tuccillo18}. In this work we are interested instead in training a network in a self-supervised way, i.e. without providing information on the true model's parameters. A general exponential profile of a galaxy observed with some inclination and position angle can be expressed as:

\begin{equation}\label{eq:exp}
I(r) = I_0 \exp { (-r^{\prime} )} 
\end{equation}

\noindent where the radius $r^{\prime}$ is computed as:

\begin{equation}
r^{\prime} = \sqrt{ (x^{\prime}/A)^2 + (y^{\prime}/B)^2  }.
\end{equation}

\noindent The major and minor semi-axis $A, B$ of the ellipsoid are related to the ellipticity as  $e = 1 - B/A$. The rotated coordinates $x^{\prime} $ and $y^{\prime} $ are given by:

\begin{align}
x^{\prime} & = \cos{\theta} (x - x_0) - \sin{\theta} (y - y_0) \\
y^{\prime} & = \sin{\theta} (x - x_0) + \cos{\theta} (y - y_0),
\end{align}

\noindent where $\theta$ corresponds to the position angle of the ellipsoid. For simplicity we set in our analysis $I_0 = 1$ and $x_0, y_0 = 0$.

%
\subsection{Network Implementation}

The full model was implemented using the Keras library \citep{keras} with the tensorflow backend \citep{tensorflow}. Keras offers a clean API for boilerplate code such as setting up a convolutional layer while allowing direct access to tensorflow for non standard tasks.

Figure \ref{fig:implementation} shows an overview of the architecture proposed here. In practice  this particular network has three inputs: the image we wish to encode and two 2D arrays containing $x$ and $y$ coordinates used to evaluate the 2D exponential profile. Coordinates could be computed on the fly using the eager programming mode in tensorflow 2.0. However, passing the pre-computed coordinates may be computationally more efficient at the expense of increased memory consumption. For the application presented here this is not an issue.

%
\subsubsection{Encoder}

The encoder consists of 3 \textit{convolutional blocks} containing each two convolutional layers with rectified linear unit (ReLU) activations followed by a $2 \times 2$ max-pooling operation. The number of kernels in the layers of each of the three convolutional blocks is $32,64$ and 128 respectively.  
The output of the last convolutional block has 128 channels of  $8 \times 8$  pixel activation images. This layer is flattened to a \num[group-separator={,}]{8192} vector which is connected to three dense layers with sigmoid activations containing 64, 32  and 3 neurons respectively. The last three neurons output the values of the predicted parameters $\hat{A}, \hat{e}, \hat{\theta}$ (one neuron per parameter).
We use dense layers as a convenient way to generate any desired number of output parameters. 
The encoder contains all the trainable parameters in the network with \num[group-separator={,}]{812963} parameters of which \num[group-separator={,}]{526531} correspond to the dense network.

The above architecture was chosen after performing a basic manual hyperparameter exploration in order to find the smallest network that would be able to accurately encode the input images. Larger networks produce lower errors at at higher computational cost.

\begin{figure*}
  \centering
  \includegraphics[width=0.95\textwidth,angle=0.0]{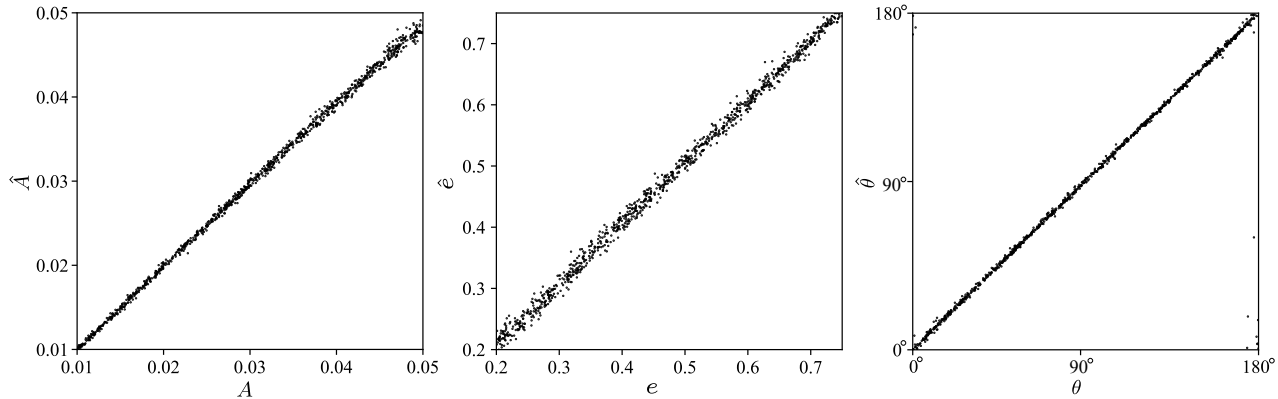}
  \caption{True vs. predicted major semi-axis $A$, ellipticity $e$ and position angle $\theta$ (left, center and right panels respectively) computed from \num[group-separator={,}]{1000} test images. The dots correspond to individual predictions. Note the failed predictions near the discontinuous interval in position angle $0^{\circ} \sim \theta \sim 180^{\circ}$.}\label{fig:comparison}
\end{figure*}

%
\subsubsection{Decoder}

The output of the encoder, the parameters $\hat{A}, \hat{e}$ and $\hat{\theta}$, are evaluated by the 2D exponential model in eq. \ref{eq:exp} acting as a decoder. In order to ``expand'' the model's parameters into a 2D image we evaluate the exponential function with the predicted parameters $\hat{A}, \hat{e}, \hat{\theta}$, and the two auxiliary arrays (of the same size as the original image) containing the $x$ and $y$ coordinates with origin at the center of the image and running in the range (-1,1). 
The decoder was implemented as a custom tensorflow layer with three inputs: the list of parameters ($\hat{A},\hat{e},\hat{\theta}$) coming from the encoder and the 2D arrays containing the coordinates $x$ and $y$ (see Fig. \ref{fig:implementation}). The output of the decoder layer is the reconstructed image $\hat{\bm{x}}$. The exponential profile in eq. \ref{eq:exp} was implemented using native tensorflow functions in order to allow tensorflow to automatically compute derivatives during backpropagation at training.

%
\subsubsection{Training}

Training data consisted of a set of \num[group-separator={,}]{10000} synthetic images of exponential profiles with a resolution of $64 \times 64$ pixels. The parameters $A,e, \theta$ of each image were sampled from uniform distributions in the ranges $0.01 < A < 0.05$, $0.2 < e < 0.75$ and $0^{\circ} < \theta < 180^{\circ}$. The position angle $\theta$ was normalized in the range $(0,1)$ to bring it inside the range of the sigmoid functions used as activations in the neurons leading to the bottleneck. In order to have a more realistic setting and to test the denoising properties of the semantic autoencoder we added Gaussian noise with dispersion $\sigma_G = 0.05$ to each image.

Training was done by feeding the model with images in batches of $20$ samples for \num[group-separator={,}]{1000} epochs. Weights were updated using the {\tiny ADADELTA} optimizer \citep{Zeiler12} with the default values of initial learning rate $l_r = 1$ and Adadelta decay factor $\rho=0.95$. We used the mean absolute error ({\it mae}) between the input and reconstructed images as the loss function. The {\it mae} produces losses that are less sensitive to the high dynamic range of the exponential profiles compared to the more commonly used mean squared error. Training took 3.9 hours on an nvidia 1050 GPU card.

%
\subsection{Results}

After training the network we tested it with \num[group-separator={,}]{1000} images not included in the training process. Prediction took a few seconds. The network produces two outputs: the reconstructed 2D profile and the model's parameters at the network's bottleneck. The parameters were extracted from the network as the activations at the last three neurons in the decoder. Figure \ref{fig:comparison} shows the comparison between the true and predicted parameters. The three parameters are recovered with excellent agreement. The network had difficulty predicting position angles close to $0^{\circ}$ and $180^{\circ}$ ($\sim 1$ percent of the images) due to the discontinuous nature of the position angle. The dispersion of the errors between the true and predicted parameters is $\sigma_A = 0.00055$, $\sigma_e = 0.0091$ and $\sigma_{\theta}=0.92^{\circ}$.  We computed  $\sigma_{\theta}$ using only points in the interval $10^{\circ}<\theta<170^{\circ}$ in order to avoid the ambiguous predictions at angles close to $0^{\circ}$ and $180^{\circ}$

\begin{figure}
  \centering
  \includegraphics[width=0.49\textwidth,angle=0.0]{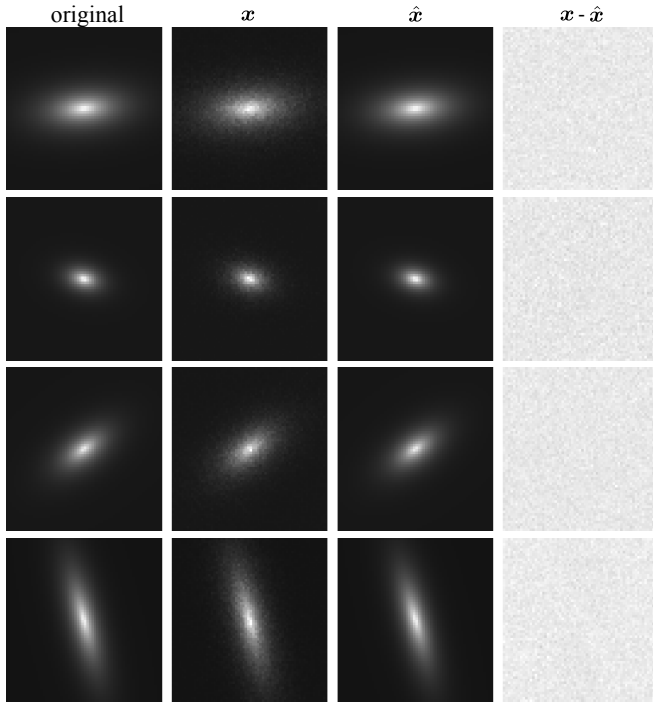}%
    \caption{From left to right:  (original) images, original images with added Gaussian noise (${\bm x}$),  reconstructed output $\hat{{\bm x}}$, and residual images (${\bm x} - \hat{{\bm x}}$) for four random combinations of the parameters $A, e, \theta$. The original images are shown here only for comparison. Both training and testing were done using the images with added Gaussian noise ${\bm x}$.}\label{fig:tri_images}
\end{figure}

Figure \ref{fig:tri_images} shows a comparison between images feed to the semantic autoencoder and the images reconstructed from the predicted parameters $A, e, \theta$. Visually, the original images without noise are indistinguishable from the reconstructed images as already indicated by the tight correlation between true and predicted parameters in Fig. \ref{fig:comparison}. The residual of the reconstructed and the input image ($\hat{{\bm x}} - {\bm x} $) shows only the Gaussian noise applied to the original images. In a sense, the semantic autoencoder presented here acts as an ideal denoising autoencoder provided that the model $G$ encodes the relevant features in the input images. 


%
\section{Discussion and future work}

We presented Semantic Autoencoders, a hybrid Deep Learning architecture that combines standard neural network elements with an explicit model describing the measured input data. We applied this architecture to the problem of self-supervised 2D exponential profile fitting with excellent results. The network is able to accurately estimate the parameters of the model with the only exception of position angles close to the ambiguous case of $0^{\circ} \sim \theta \sim 180^{\circ}$. This is not a limitation of the architecture presented here but the result of the network trying to interpolate in the discontinuous interval.

Semantic autoencoders act as ideal denoising autoencoders, in the sense that they reconstruct the input image as a function of the features we consider most relevant. As such, their efficacy fully depends on the model used as decoder. This ``Bayesian'' twist on the standard autoencoder can be exploited for model testing, assuming that the encoder is sufficiently complex to learn to estimate the model's parameters.


In general, problems that can be solved with Semantic Autoencoders can also be solved with standard recursive optimization methods. What the proposed architecture offers is a model that can solve, in an unsupervised way, for all the possible cases (within the region in feature space covered by the training sample) in a fast and efficient way. A machine learning approach trades network complexity and  long training time for robustness and speed at prediction time.

The work presented here highlights the possibility of constructing of more general neural networks that combine standard artificial neurons with general algorithms or even physical systems. The decoder function $G$ could be any process that is differentiable including analytic functions, numerical algorithms \citep{Li18} and even real-world physical systems. 

While this work was being independently developed google released the {\tiny TENSORFLOW GRAPHICS} deep learning framework \citep{tensorflowgraphics} which is an implementation of the basic architecture presented here applied to computer graphics.

%
\section{Acknowledgments}

The author would like to thank Mark Neyrinck for useful comments on the manuscript. This work was funded  in part by ``Programa de Apoyo a Proyectos de Investigaci\'{o}n e Innovaci\'{o}n Tecnol\'{o}gica'' grant DGAPA-PAPIIT IA104818.



\begin{thebibliography}{}

\bibitem[Abadi, M. et al., (2015)]{tensorflow}
Abadi, M. et al., 2015, TensorFlow: Large-scale machine learning on heterogeneous systems, tensorflow.org.

\bibitem[\protect\citeauthoryear{Barden, H{\"a}u{\ss}ler, Peng, McIntosh \& Guo}{2012}]{Barden12} 
Barden M., H{\"a}u{\ss}ler B., Peng C.~Y., McIntosh D.~H., Guo Y., 2012, MNRAS, 422, 449

\bibitem[Bischof C. H., (1996)]{Bischof96}
Bischof C. H., 1996, IUTAM Symposium on Optimization of Mechanical Systems, Springer Netherlands, 41-48.

\bibitem[Bourlard H., Kamp Y., (1988)]{Bourlard88}
Bourlard H., Kamp Y., 1988, Biol. Cybern., 59, 291 

\bibitem[Chollet F. et al., (2015)]{keras}
Chollet F. et al., 2015, https://github.com/fchollet/keras

\bibitem[Cybenko G., (1989)]{Cybenko89}
Cybenko G., 1989, MCSS, 2, 303

\bibitem[\protect\citeauthoryear{Dieleman, Willett \& Dambre}{2015}]{Dieleman15} 
Dieleman S., Willett K.~W., Dambre J., 2015, MNRAS, 450, 1441

\bibitem[Hinton G., Salakhutdinov R., (2006)]{Hinton06}
Hinton G., Salakhutdinov R., 2006, Science (New York, N.Y.), 313, 504

\bibitem[Dom{\'\i}nguez S{\'a}nchez, Huertas-Company, Bernardi, Tuccillo \& Fischer (2018)]{Sanchez18}
Dom{\'\i}nguez S{\'a}nchez H., Huertas-Company M., Bernardi M., Tuccillo D., Fischer J.~L., 2018, MNRAS, 476, 3661

\bibitem[Iten R., et al., (2018)]{Iten18}
Iten R., Metger T., Wilming H., del Rio L., Renner R., 2018, arXiv e-prints, arXiv:1807.10300

\bibitem[\protect\citeauthoryear{Kim \& Brunner}{2017}]{Kim17} 
Kim E.~J., Brunner R.~J., 2017, MNRAS, 464, 4463

\bibitem[LeCun Y., (1987)]{LeCun87}
LeCun Y., 1987, PhD thesis, Universite de Paris VI

\bibitem[LeCun Y., Bengio Y., Hilton G.(2015)]{LeCun15}
LeCun Y., Bengio Y., Hilton G., 2015, Nature, 521,436

\bibitem[Legendre A. M., (1805)]{Legendre}
Legendre A. M., 1805, Nouvelles m\'{e}thodes pour la d\'{e}termination des orbites des cometes,  Sur la M\'{e}thode des moindres quarr\'{e}s, Paris: F. Didot.

\bibitem[Li T.-M. et al. (2018)]{Li18}
Li T.-M., Aittala M., Durand F., Lehtinen J., 2018, ACM Trans. Graph. (Proc. SIGGRAPH Asia), 37, 222:1

\bibitem[Peng C. et al. (2002)]{Peng02}
Peng C. Y., Ho L. C., Impey C. D., Rix H.-W., 2002, A$\&$A, 124, 266

\bibitem[S{\'e}rsic (1963)]{Sersic63} 
S{\'e}rsic, J.~L.\ 1963, Boletin de la Asociacion Argentina de Astronomia La Plata Argentina, 6, 41 

\bibitem[\protect\citeauthoryear{Simard, Mendel, Patton, Ellison \& McConnachie}{2011}]{Simard11} Simard L., Mendel J.~T., Patton D.~R., Ellison S.~L., McConnachie A.~W., 2011, ApJS, 196, 11

\bibitem[Tuccillo D. et al. (2018)]{Tuccillo18}
Tuccillo D., Huertas-Company M., Decenci`ere E., Velasco-Forero S., Domınguez Sanchez H., Dimauro P., 2018, MNRAS, 475, 894

\bibitem[Valentin, J., et al. (2019)]{tensorflowgraphics}
Valentin, J., Keskin, C., Pidlypenskyi, P., Makadia A.,  Sud, A., and Bouaziz, Sofien, 2019, TensorFlow Graphics. https://github.com/tensorflow/graphics.

\bibitem[Vincent P. et al. (2010)]{Vincent10}
Vincent P., Larochelle H., Lajoie I., Bengio Y., Manzagol P.-A., 2010, J. Mach. Learn. Res., 11, 3371

\bibitem[Wengert, R. E, (1964)]{Wengert64}
Wengert, R. E, 1964, Communications of the ACM., 7, 463-464

\bibitem[\protect\citeauthoryear{Zeiler}{2012}]{Zeiler12} 
Zeiler M.~D., 2012, arXiv, arXiv:1212.5701

\end{thebibliography}
\end{document}